\documentclass[twocolumn,epjc3]{svjour3}

\RequirePackage{graphicx} 

\usepackage{dcolumn}
\usepackage{bm}
\usepackage{epsfig}
\usepackage{figlatex}
\usepackage{pdflscape}
\usepackage{subfigure} 
\usepackage{multirow}
\usepackage[colorlinks=true,citecolor=blue,filecolor=blue,linkcolor=blue,urlcolor=blue,pdftex]{hyperref}
\usepackage{dcolumn}

\usepackage[usenames,dvipsnames]{xcolor}
\usepackage[normalem]{ulem}

\journalname{Eur. Phys. J. C}

\begin{document}
\title{Cosmogenic activation of xenon and copper}

\author{Laura Baudis\thanksref{e1,addr1}
  \and Alexander Kish\thanksref{e2,addr1}
  \and Francesco Piastra\thanksref{e3,addr1}
  \and Marc Schumann\thanksref{e4,addr2}
}
\thankstext{e1}{e-mail: lbaudis@physik.uzh.ch}
\thankstext{e2}{e-mail: alexander.kish@physik.uzh.ch}
\thankstext{e3}{e-mail: fpiastra@physik.uzh.ch}
\thankstext{e4}{e-mail: marc.schumann@lhep.unibe.ch}

\institute{Department of Physics, University of Z\"{u}rich, Winterthurerstrasse 190, CH-8057, Z\"{u}rich, Switzerland \label{addr1}
\and Albert Einstein Center for Fundamental Physics, University of Bern, Sidlerstrasse 5, CH-3012 Bern, Switzerland \label{addr2} }

\date{}

\maketitle

\begin{abstract}
Rare event search experiments using liquid xenon as target and detection medium  require ultra-low background levels to fully exploit their physics potential. Cosmic ray  induced activation of the detector components and, even more importantly, of the xenon itself during production, transportation and storage at the Earth's surface, might result in the production of radioactive isotopes with long half-lives, with a possible impact on the expected background. We present the first dedicated study on the cosmogenic activation of xenon after 345~days of exposure to cosmic rays at the Jungfraujoch research station at 3470\,m above sea level, complemented by a study of copper which has been activated simultaneously. We have directly observed the production of $^7$Be, $^{101}$Rh, $^{125}$Sb, $^{126}$I and $^{127}$Xe in xenon, out of which only $^{125}$Sb could potentially lead to background for a multi-ton scale dark matter search. The production rates for five out of eight studied radioactive isotopes in copper are in agreement with the only existing dedicated activation measurement, while we observe lower rates for the remaining ones. The specific saturation activities for both samples are also compared to predictions obtained with commonly used software packages, where we observe some underpredictions, especially for xenon activation.
\end{abstract}

\section{Introduction}

Liquid xenon (LXe) is used as detection medium in current and future rare event search experiments, such as direct dark matter~\cite{XE100:2012,XMASS:2013,LUX:2013,XE1T:2013,DARWINwimp} and neutrinoless double-beta decay searches~\cite{EXO:2012,KamLAND,NEXT:2013}. It features a high scintillation and ionization yield~\cite{Plante2011,Aaron2010}, as well as a high radio-purity. Apart from long-lived double-beta emitters, such as $^{124}$Xe, $^{126}$Xe, $^{134}$Xe and $^{136}$Xe, where only the decay of  $^{136}$Xe has been observed so far, xenon has no unstable isotopes. However, the exposure to cosmic rays during production, transportation and storage aboveground can produce instable radio-isotopes via nuclear activation processes.

For a given exposure time and altitude, the activation yield of materials can be predicted using software packages such as Activia~\cite{ACTIVIA} and Cosmo~\cite{COSMO}. Calculations with both codes were performed for the XENON100 experiment~\cite{XE100:2012}, however the predicted production rates were too high to be compatible with the measured background rates~\cite{Xe100:EMBG,AlexThesis,Gondolo}.

In this work we present the first dedicated experimental measurement of the cosmogenic activation of a natural xenon sample, which we exposed to cosmic rays for 345\,days at an altitude of 3470\,m. In order to provide a benchmark for our activation measurements and predictions, a sample of oxygen-free high thermal conductivity (OFHC) copper has been simultaneously exposed at the same location. Due to its very high purity, and thus low radioactivity levels, OFHC copper is one of the most-frequently used materials to construct low-background detectors, and a well validated material in the software packages.

\begin{table*}[tb]
\setlength\extrarowheight{2pt}
\centering
\caption{Isotopic composition of natural xenon and copper~\cite{IUPAC:IsotAbund}. The half-lives of the double-beta decay isotopes $^{124}$Xe, $^{134}$Xe and $^{1236}$Xe are from the NuDat~2.6 database~\cite{NuDat}, the range for the double electron capture isotope $^{126}$Xe is a theoretical prediction from~\cite{ref::shukla}. \label{tab:XeAbound}}
\vspace{0.2cm}
\begin{tabular}{lccccccccc}
\hline
Xe mass number & 124 & 126 & 128 & 129 & 130 & 131 & 132 & 134 & 136 \\ 
Abundance [\%] & 0.09(1) & 0.09(1) & 1.92(3) & 26.44(24) & 4.08(2) & 21.18(3) & 26.89(6) & 10.44(10) & 8.87(16) \\ 
T$_{1/2}$ [y] & $>$1.6$\times$10$^{14}$ & [5-12]$\times$10$^{25}$ & -- & -- & -- & -- & -- & $>$5.8$\times$10$^{22}$ & 2.165$\times$10$^{21}$ \\ [2pt] 
\hline
Cu mass number & 63 & 65 &  &  &  &  &  &  &  \\
Abundance [\%] & 69.17(3) & 30.83(3) &  &  &  &  &  &  &  \\
\hline
\end{tabular}
\end{table*}

\begin{table*}[tbp]
\setlength\extrarowheight{2pt}
\centering
\caption{Altitude, atmospheric depth, and vertical flux of protons and neutrons, as well as scaling factors to relate the flux to sea level values for the various locations in our study. }
\label{tab:VertFlux}
\vspace{0.2cm}
\begin{tabular}{lrrrr}
\hline
Location      		& Altitude [m]   & Atm.~depth [g/cm$^{2}$]   	&  Vert.~flux [m$^{-2}$s$^{-1}$sr$^{-1}$] & Scaling factor	\\ [2pt]
\hline
Sea level 	& 0 	& 1030  & 2.6 	& 1.0		\\
Lauterbrunnen 	& 795 	& 954 	& 4.7 	& 1.8		\\
LNGS 		& 985 	& 936 	& 5.4 	& 2.1		\\
Jungfraujoch 	& 3470 	& 728 	& 29.0 	& 11.2	\\ [2pt]
\hline
\end{tabular}
\end{table*}

The intrinsic radioactivity of the samples, initially stored underground at the Laboratori Nazionali del Gran Sasso (LNGS, Italy) for more than 1.5\,y, was measured by means of a high-purity germanium $\gamma$-spectrometer at LNGS, before and after the exposure to cosmic rays. Section~\ref{Sec:Preparation} describes the samples and their handling, Section~\ref{Sec:Activation} the cosmic activation procedure and its modelling, and Section~\ref{Sec:ExpSetup} provides details on the $\gamma$-spectro\-meter and the data analysis. The results of the measurement and the comparison with predictions are presented in Section~\ref{Sec:Results}, while conclusions are drawn in Section~\ref{Sec:Conclusions}.

\section{Samples and preparation\label{Sec:Preparation}}

The xenon sample consisted of 2.04~kg research-grade xenon (impurities $<$10\,ppm) from Praxair with natural isotopic composition (see Table~\ref{tab:XeAbound}). It was contained in a 1\,liter stainless steel bottle at a pressure of $\sim$100\,bar (at $\overline{T} \sim 20^\circ$C). To avoid the measurement results to be dominated by activation products in the bottle ($m=2.9$\,kg), two identical bottles were used for the $\gamma$-screening and for the activation procedure. The first one always remained underground, where the gas transfer took place as well. Both bottles were evacuated and baked at $>$100$^\circ$C for several days before being filled with xenon.  

The 10.35\,kg copper sample consisted of 5~blocks of OFHC copper (purity 99.99\%, isotopic composition in Table~\ref{tab:XeAbound}) from Norddeutsche Affinerie (now Aurubis). It came from the batch used to construct inner parts of the XENON100 detector (sample~6 in~\cite{XE100:screen}). Before each measurement, surface contaminations were removed in an ultrasonic bath filled with the acid detergent Elma Clean~60 diluted with deionized water. The sample was then rinsed and wiped with pure ethanol ($>98$\%), and stored under boil-off N$_{2}$ atmosphere for several days in order to let the $^{222}$Rn diffuse out and decay.

\section{Activation by cosmic rays \label{Sec:Activation}}

The activation took place in a controlled manner at the High Altitude Research Station Jungfraujoch~\cite{Jungfrau}, at an altitude of 3470\,m above sea level, from October 31$\mathrm{^{st}}$, 2012 to October 15$\mathrm{^{th}}$, 2013. The cumulated activation time was 345.0~days. The initial transport from the underground laboratory of LNGS to the Jungfraujoch took less than 5~days and is neglected in the analysis. Following the activation, the samples were brought to Lauterbrunnen (795\,m), where they were stored for about 4~days before being transported by car to LNGS for the underground $\gamma$-measurement. Transport to and storage at the LNGS aboveground laboratory (985\,m) lasted about 1~day. The cool-down times, from the time when the samples were brought underground until the start of the measurement, were 2.5~days and 14.8~days for the xenon and copper samples, respectively.

\begin{figure*}[ht]
\centering
\subfigure[]{
\includegraphics[width=0.3\linewidth]{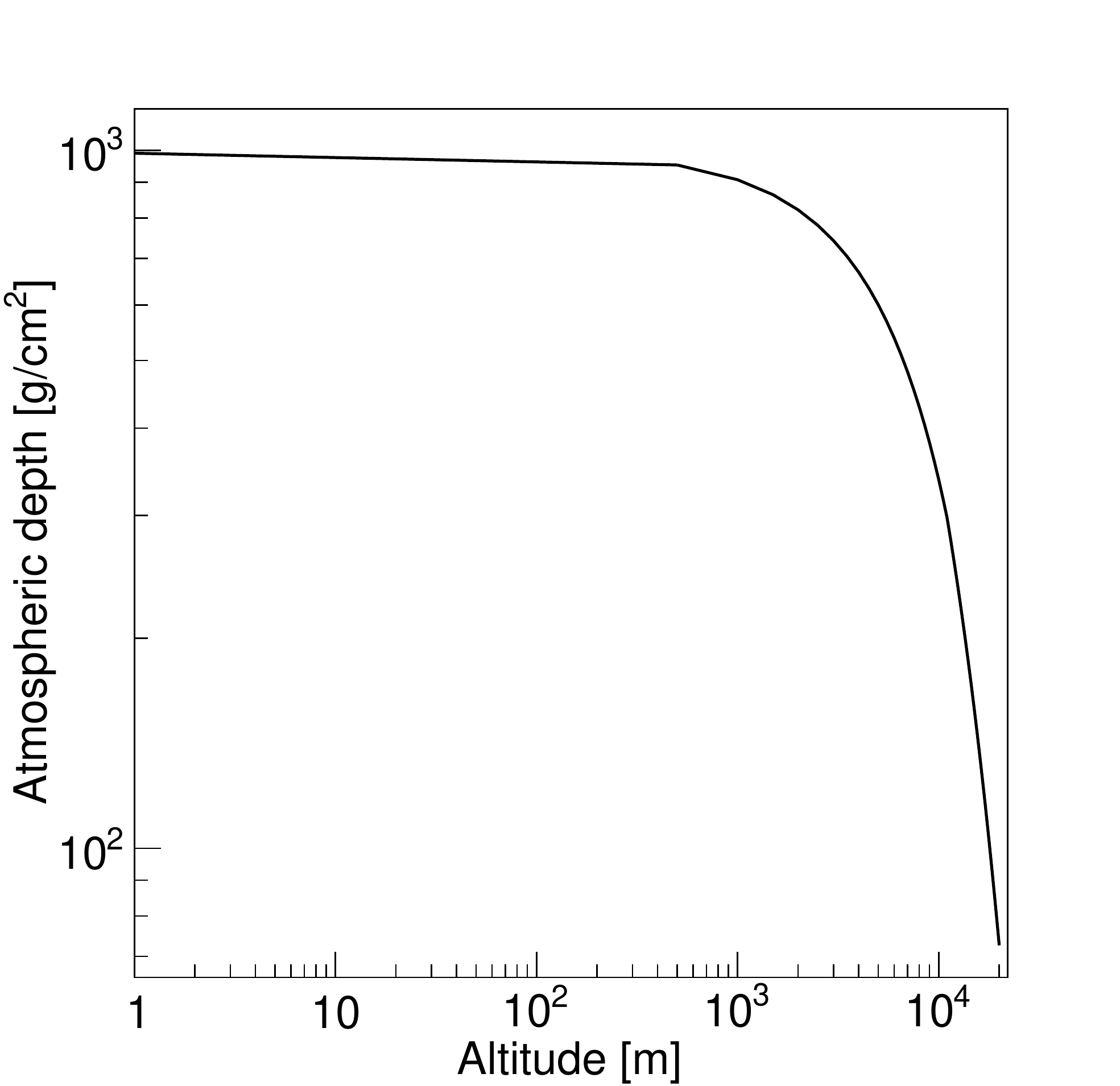}
\label{fig:CosmicRays1}}
\subfigure[]{
\includegraphics[width=0.3\linewidth]{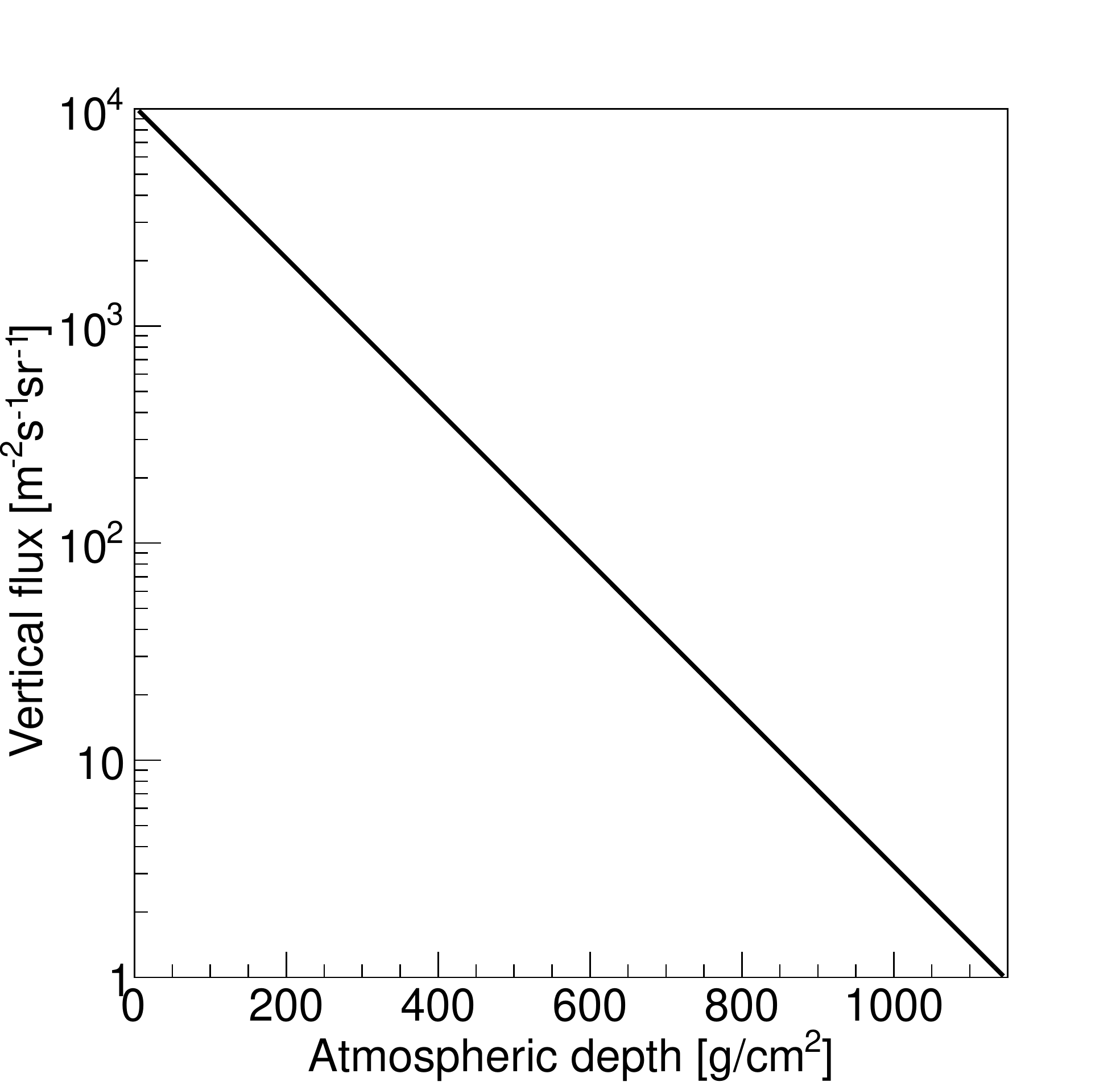}
\label{fig:CosmicRays2}}
\subfigure[]{
\includegraphics[width=0.3\linewidth]{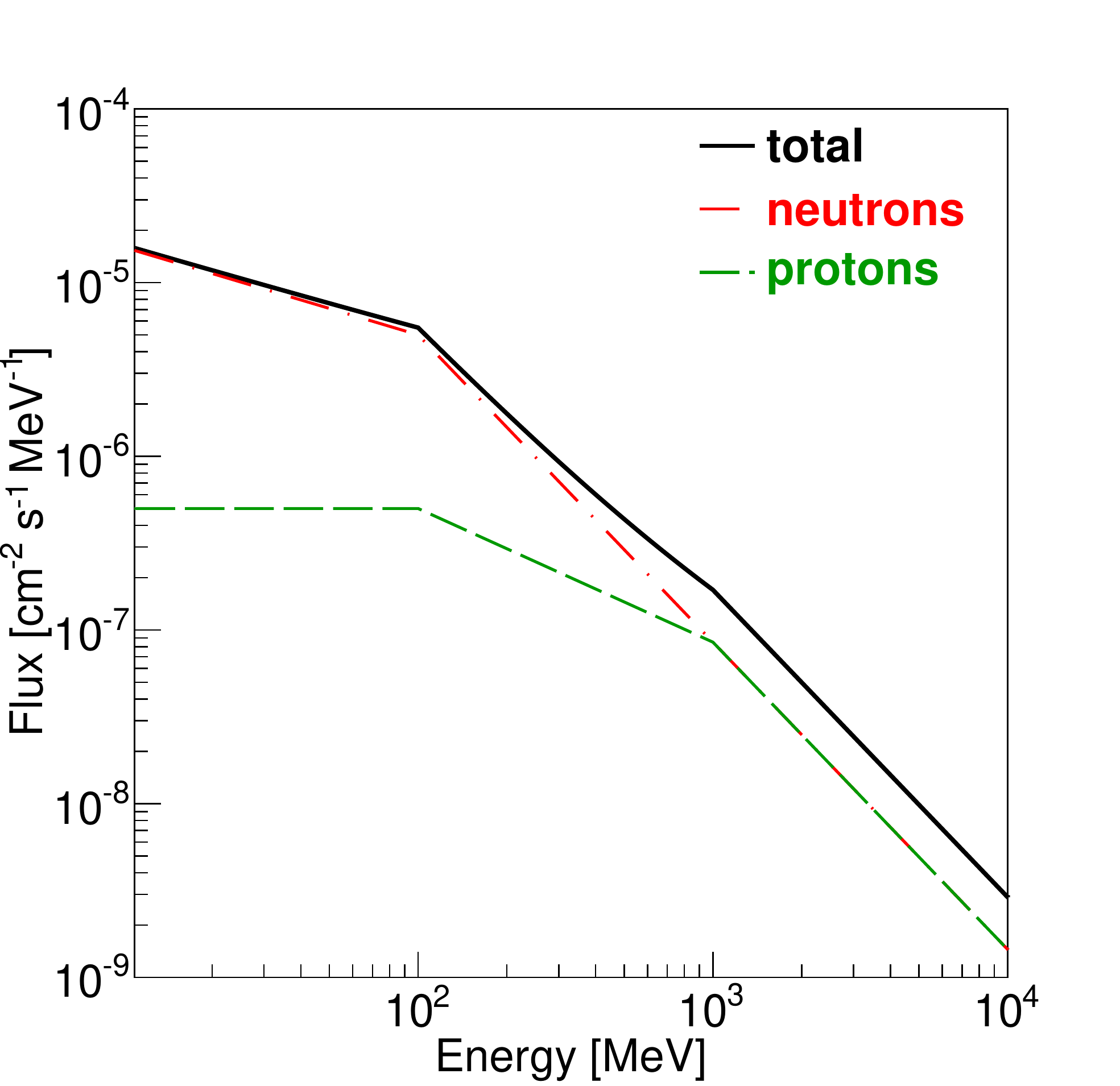}
\label{fig:CosmicRays3}}
\caption{(a) Relation between atmospheric depth and altitude from the U.S.~Standard Atmosphere Model~\cite{AtmModel}. We adopt a sea level value of 1030~g/cm$^{2}$~\cite{AtmDepth_Ziegler,AtmDepth_Stanev}. (b) Vertical nucleon flux as a function of the atmospheric depth~\cite{PDG_VertFlux}. (c) The cosmic ray spectrum from Activia and Cosmo, based on parametrization from Refs.~\cite{CRspectrum_Armstrong,CRspectrum_Gehrels}. Only the total flux is used in the calculations as protons and neutrons interact similarly at these energies.  \label{fig:CosmicRays}}
\end{figure*}

For comparison with theoretical predictions and with other measurements, we have to convert our high-alti\-tude activation results into specific saturation activities at sea level. We thus calculate scaling factors that relate the nucleon cosmic ray flux at the various locations, as shown in Table~\ref{tab:VertFlux}.  The atmospheric depth at different altitudes was obtained using the U.S. Standard Atmosphere 1976 model~\cite{AtmModel}, shown in Figure~\ref{fig:CosmicRays1}, with a sea level value of 1030\,g/cm$^{2}$ \cite{AtmDepth_Ziegler,AtmDepth_Stanev}. The vertical flux of protons and neutrons was then calculated for a given atmospheric depth based on Ref.~\cite{PDG_VertFlux}, see Figure~\ref{fig:CosmicRays2}. Due to their lower flux~\cite{ref::Beatty} and lower cross section~\cite{ref::Heisinger,ref::Hagner} muons only play a sub-dominant role in activation processes and are neglected here.

The cosmogenic activation was predicted using Activia~\cite{ACTIVIA} and an updated and revised version of Cosmo \cite{COSMO,Martoff}. The cosmic ray spectrum encoded in the programs is based on the parametrization from Refs.~\cite{CRspectrum_Armstrong,CRspectrum_Gehrels} and is shown in Figure~\ref{fig:CosmicRays3}. It was sampled from 10\,MeV to 10\,GeV in 10\,MeV intervals: the lower boundary is chosen because the energy thresholds of the nuclear excitation functions are above this energy. The cosmic ray flux above 10\,GeV does not affect our results due to the exponential behavior of the energy spectrum. Both packages calculate the cross-sections of the relevant nuclear processes, such as spallation, fission, and evaporation, using semi-empirical formulae developed by Silberberg and Tsao~\cite{SilberbergTsao}, with identical parameters for protons and neutrons. To distinguish between spallation, fission, and nuclear breakdown reactions, including contributions from fast fragmentation processes and transition areas, these formulae are defined in separate regions, depending on the masses of the target and product nuclides (light, intermediate and heavy isotopes). Above a nucleon energy of $\sim$3~GeV, the cross-sections are assumed to be energy-independent~\cite{ACTIVIA,SilberbergTsao}. 

Our calculations of the production rates were performed for an exposure to cosmic rays at sea level. To convert into saturation activities, we assumed an activation time of 100\,y and no cool-down time.

\section{Measurement and data analysis \label{Sec:ExpSetup}}

Before and after activation, the intrinsic radioactivity of the xenon and copper samples was measured with the high-purity germanium $\gamma$-spectrometer Gator~\cite{GatorPaper}, operated underground at LNGS. The detector has a background rate of 230\,counts/day in the 100-2700\,keV interval. Due to the $\sim$3500\,m water equivalent of rock shielding from cosmic radiation, the hadronic component of the cosmic radiation is completely absent and the atmospheric muon flux is suppressed by 6~orders of magnitude with respect to the aboveground laboratory~\cite{MACRO_1}. Hence it is safe to assume that no significant cosmogenic activation took place once the samples were stored underground.

\begin{figure*}[ht]
\includegraphics[width=1.0\textwidth]{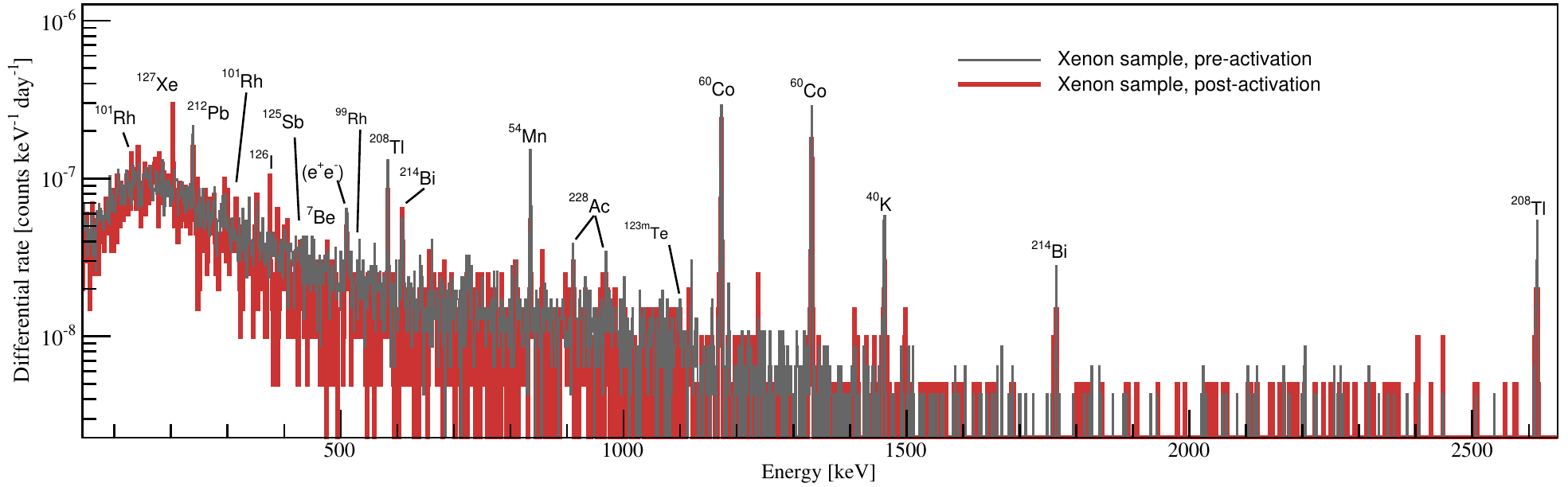}
\caption{Pre- and post-activation spectra of the 2.04\,kg xenon sample. One can identify the $^{127}$Xe lines at 202.9\,keV and  375.0\,keV, and the $^{126}$I line at 388.6\,keV. Other prominent lines are from radioactive contaminations in the stainless steel bottle containing the xenon (primordial $^{238}$U and $^{232}$Th chains, $^{40}$K, cosmogenic $^{54}$Mn, $^{60}$Co).\label{fig:SpectraXe}}
\end{figure*}

\begin{figure*}[ht]
\includegraphics[width=1.0\textwidth]{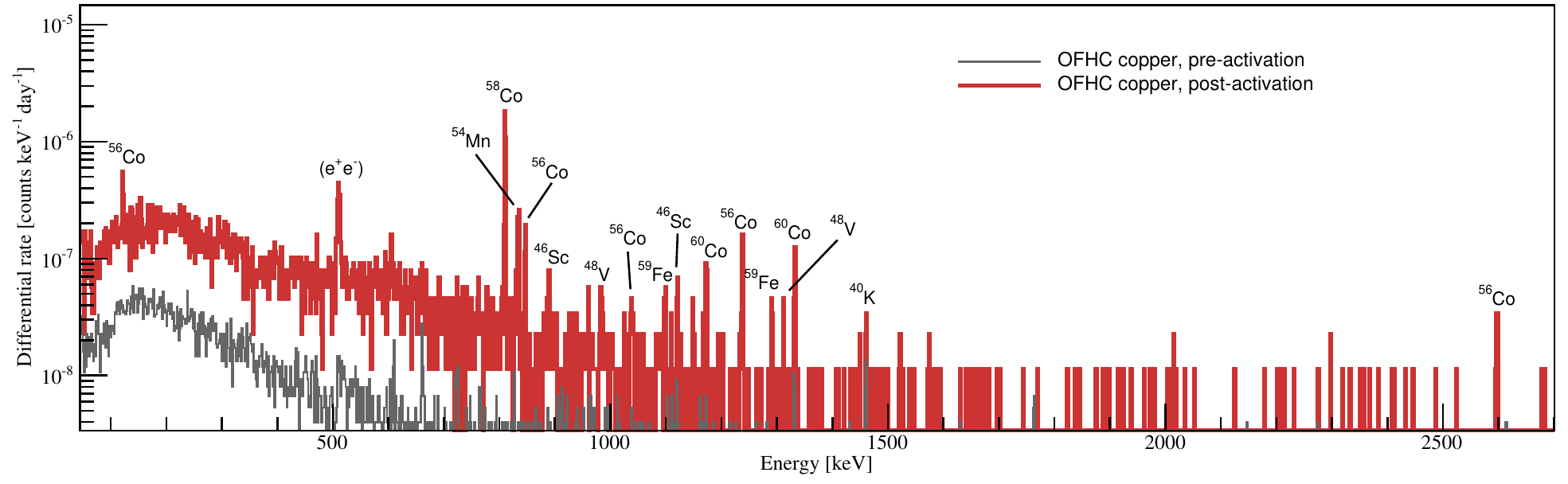}
\caption{Pre- and post-activation spectra of the 10.35\,kg OFHC copper sample. The entire post-activation spectrum is dominated by the cosmogenic activation products. \label{fig:SpectraCu}}
\end{figure*}

To establish the detection sensitivity to the $\gamma$-lines of the expected activation products as a function of measuring time, the xenon and the copper samples were measured pre-activation for 26.5~days and 34.3~days, respectively. The post-activation spectra were acquired for 11.5~days for xenon and for 4.0~days for copper. The specific activities of the produced radio-isotopes are inferred from the intensities of their most prominent full absorption peaks in the post-activation spectra. The information in the pre-activation measurements is not used in the analysis as no cosmogenic isotopes were present. Both, pre-activation and post-activation spectra are shown in Figures~\ref{fig:SpectraXe} and~\ref{fig:SpectraCu}.

A Bayesian method, described in Refs.~\cite{Weise1989,Weise1992,Lira2010}, was employed to infer the activity of an isotope using all its $\gamma$-lines with a sufficiently large branching ratio. The spectrum around each line is divided into three regions, their width is related to the energy-dependent resolution $\sigma$ of the detector which is given in~\cite{GatorPaper}. The signal region is defined as $\pm$3$\sigma$ around the mean position of the full absorption peak. The background is inferred from the count rates in two control regions, +3$\sigma$ above and $-$3$\sigma$ below the signal region. The total count rate $\gamma_S$ in the signal region is given by 
\begin{equation}
\gamma_{\mathrm{S}} = m\cdot A (\varepsilon\cdot \textnormal{BR}) +w_{S}\frac{\beta_{\mathrm{L}}+\beta_{\mathrm{R}}}{ w_{\mathrm{L}}+w_{\mathrm{R}} },
\end{equation}
where $m$ is the sample mass, $A$ is the specific activity of the sample, $\beta_{\mathrm{L,R}}$ are the background rates in the left and right control regions, respectively, and $w_{\mathrm{S,L,R}}$ are the widths of the three regions. Hence the background rate in the signal region is the interpolation of the background rates in the control regions. The product of branching ratio~BR and detection efficiency~$\varepsilon$ of the $\gamma$-line is calculated in Monte Carlo simulations, based on a detailed implementation of the detector and sample geometry in GEANT4~\cite{geant4}. The decays of the radio-isotopes of interest are simulated using the G4Radioactive\-Decay class, where branching ratios and directional correlations are taken into account.

Considering $N$ $\gamma$-lines for the same isotope, the likelihood of the model is
\begin{equation}
\mathcal{L} = \prod\limits_{k}^{N} f_P(C_{ \mathrm{S}_{k} }|\gamma_{\mathrm{S}_{k}}t)\cdot f_P(C_{ \mathrm{L}_{k} }|\beta_{ \mathrm{L}_{k} }t)\cdot f_P(C_{ \mathrm{R}_{k} }|\beta_{ \mathrm{R}_{k} }t),
\end{equation}
where the $C_{ \mathrm{S}_{k}\mathrm{,L}_{k}\mathrm{,R}_{k}}$ are the counts in the three regions for each line $k$, and $t$ is the measurement time. $f_P(C|\mu)=\mu^C/C! \ e^{-\mu}$ is the Poisson distribution.

\begin{figure*}[ht]
\centering
\subfigure[]{
\includegraphics[width=0.3\linewidth]{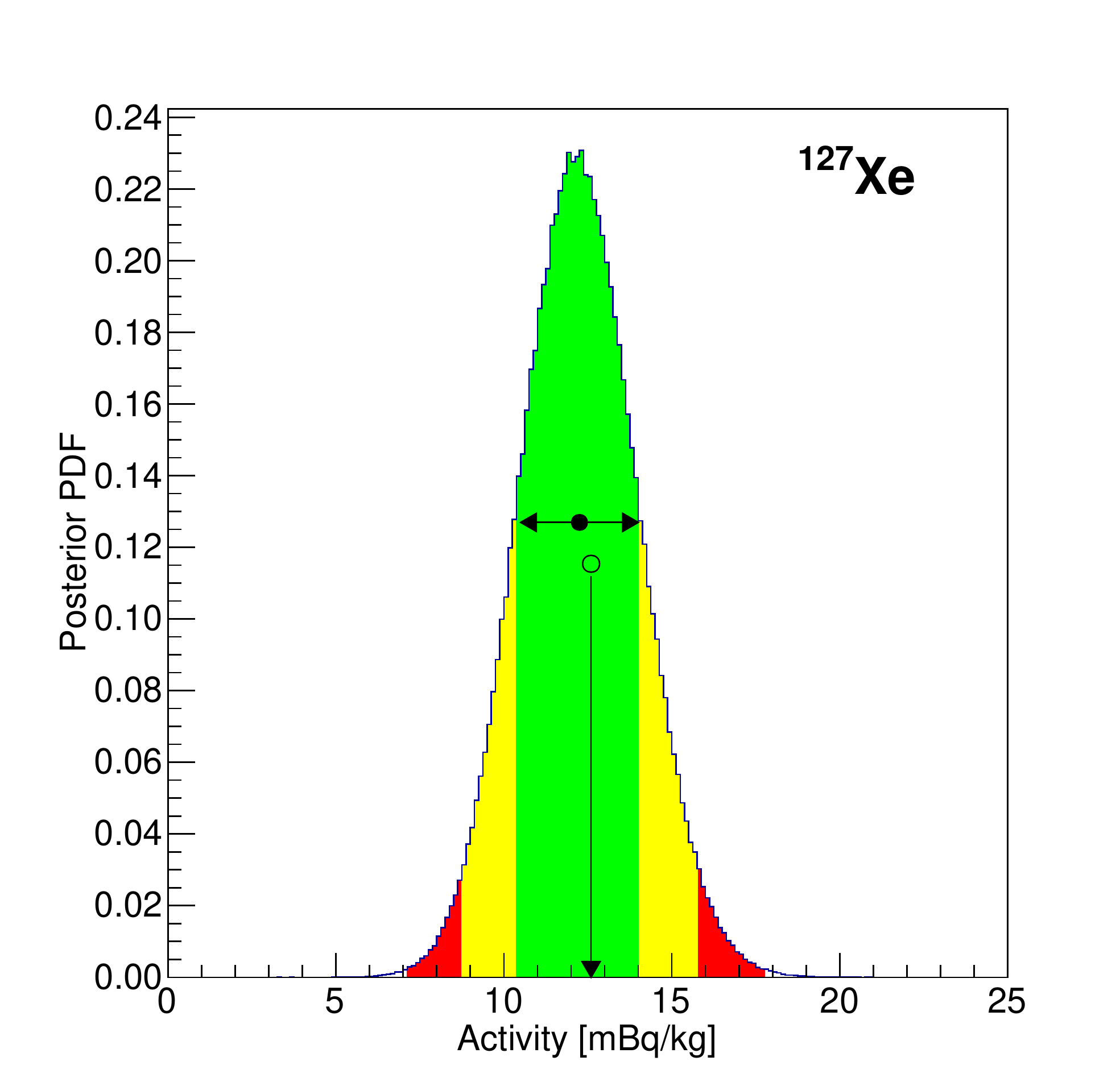}
\label{fig:posteriors1}}
\subfigure[]{
\includegraphics[width=0.3\linewidth]{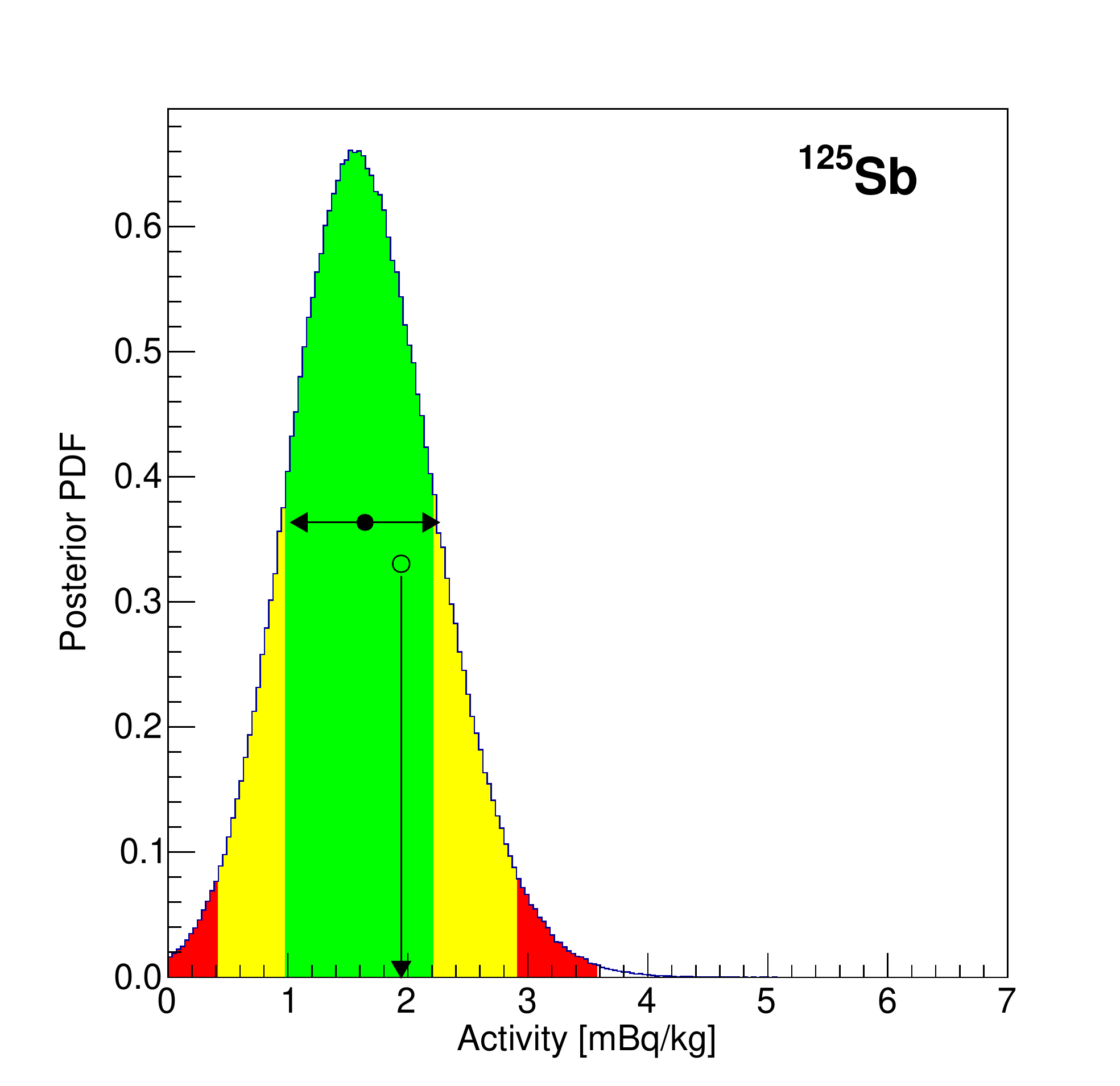}
\label{fig:posteriors11}}
\subfigure[]{
\includegraphics[width=0.3\linewidth]{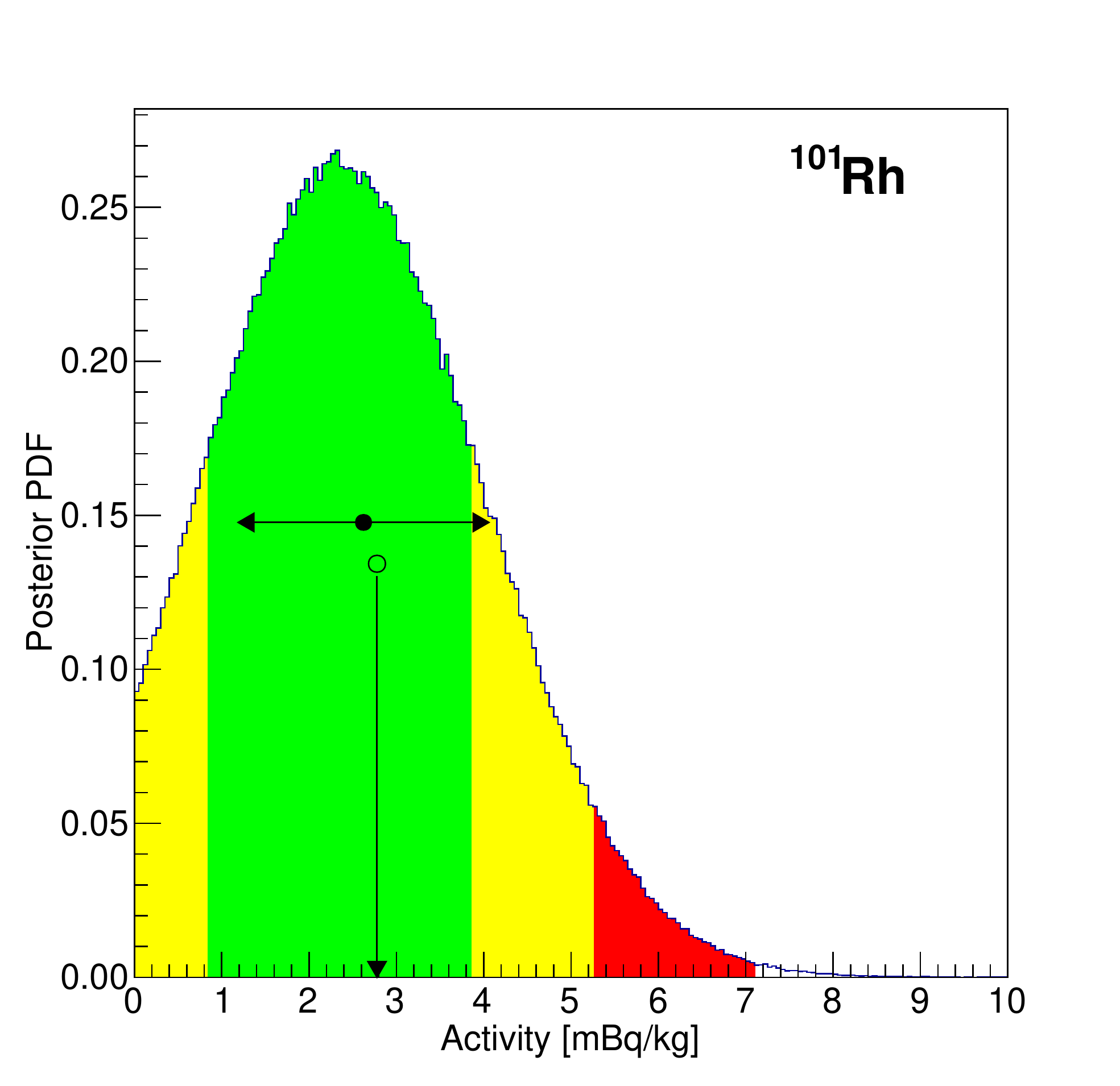}
\label{fig:posteriors111}}

\vspace{-0.4cm}
\subfigure[]{
\includegraphics[width=0.3\linewidth]{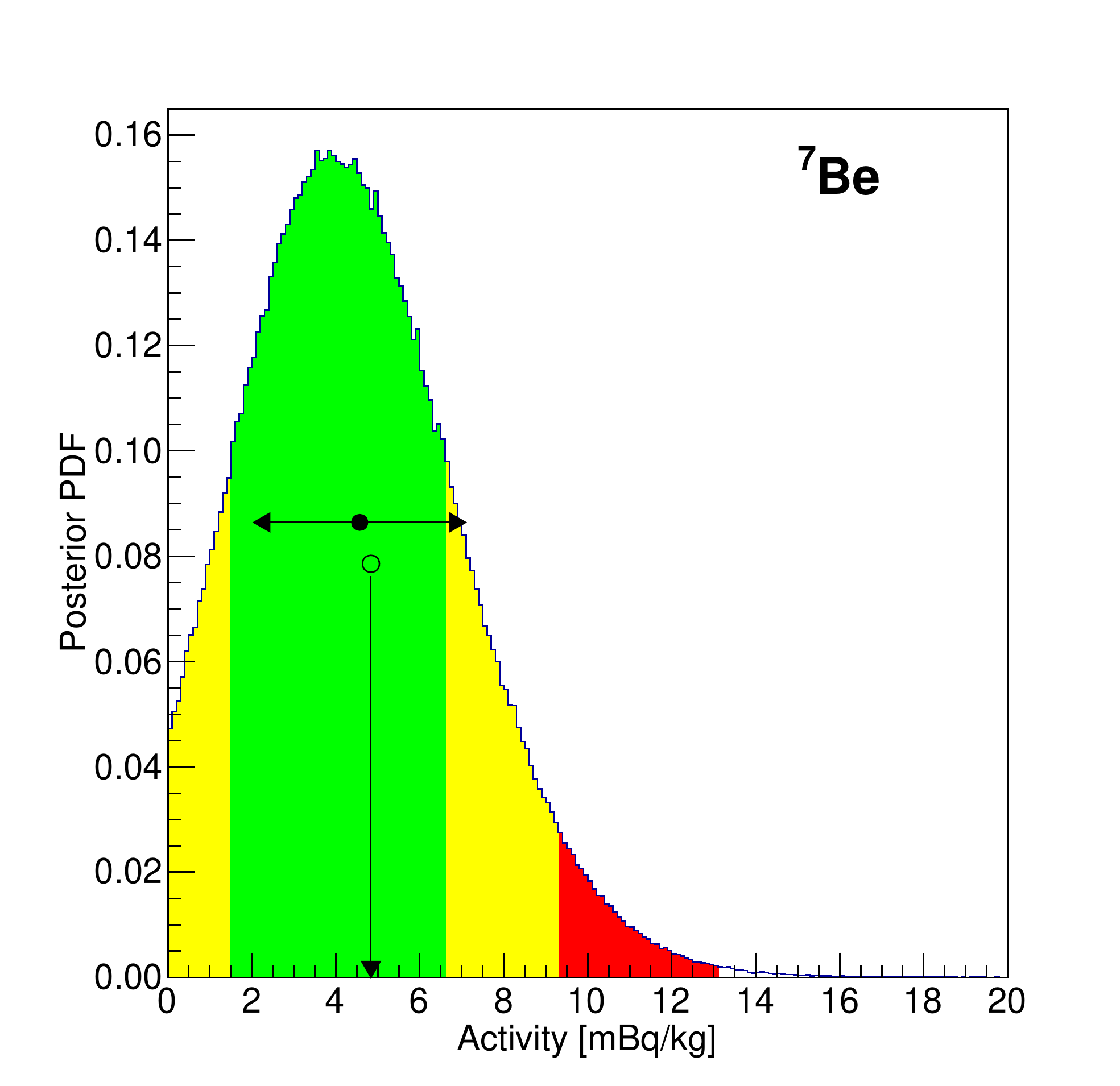}
\label{fig:posteriors22}}
\subfigure[]{
\includegraphics[width=0.3\linewidth]{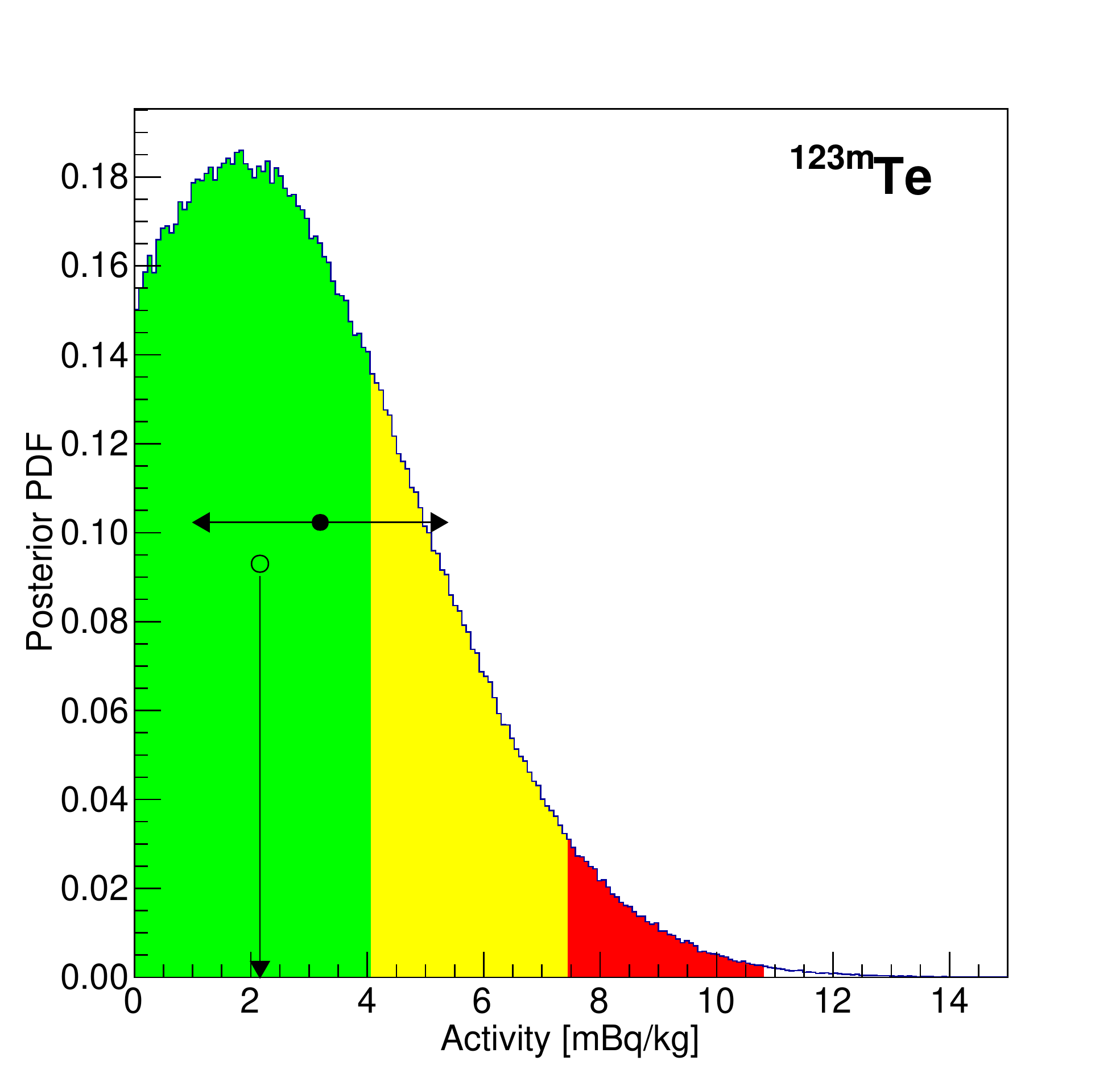}
\label{fig:posteriors2}}
\subfigure[]{
\includegraphics[width=0.3\linewidth]{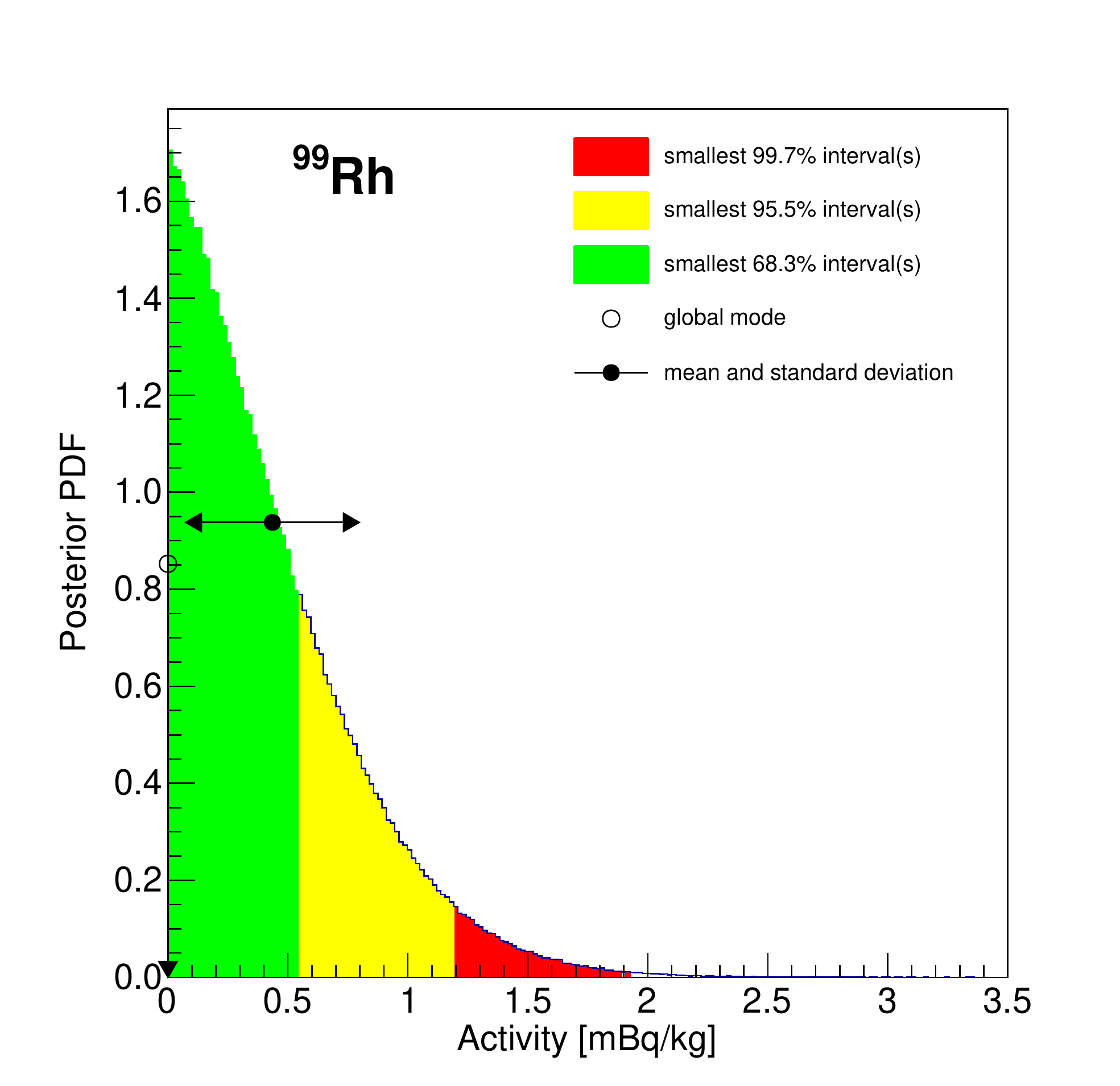}
\label{fig:posteriors3}}
\caption{(a) Posterior PDF for the detected $^{127}$Xe activity, clearly visible in Figure~\ref{fig:SpectraXe}. The green region represents the shortest 68.3\%~C.I., which is taken as $\pm$1$\sigma$ uncertainty. (b) $^{125}$Sb, (c) $^{101}$Rh and (d) $^{7}$Be are detected as well, but with larger uncertainties. (e) For $^{123\mathrm{m}}$Te a signal is present, but an upper limit is reported since the intensity is too low to be determined. (f) An upper limit is reported for the activity of $^{99}$Rh. The right edge of the yellow region represents the 95.5\%~upper limit (or the lower 95.5\% C.I.). }
\label{fig:posteriors}
\end{figure*}

We use flat priors for the parameters $\gamma_{k}$ and $\beta_{\mathrm{L}_{k}\mathrm{,R}_{k}}$ and generate the posterior probability density function (PDF) of the specific activity~$A$ by Markov Chain Monte Carlo methods implemented in the Bayesian Analysis Toolkit (BAT)~\cite{Caldwell2009}. The marginalised posterior PDF is used to decide whether we can claim the detection of a line: if the global mode of the posterior PDF and the left edge of the shortest 68.3\%~credibility interval (C.I., green region in Figures~\ref{fig:posteriors1}--\ref{fig:posteriors22}) are positive, we calculate an activity, using the mode of the posterior PDF as its estimator and the shortest 68.3\%~C.I.~as $\pm$1$\sigma$~uncertainty. If only the global mode is positive, but the left edge of the shortest 68.3\% C.I.~is zero, we report an upper limit as the signal is too weak to be determined. The upper limit is given as the 95.5\%~quantile of the posterior PDF (the right edge of the yellow region in Figures~\ref{fig:posteriors2} and~\ref{fig:posteriors3}).

Because the half-lives of several examined isotopes are comparable to the integral time of the measurement, the measured mean specific activity $A$ can significantly differ from the one at the beginning of the measurement
\begin{equation}
 A_{0} = A\cdot\left(\frac{t}{\tau}\right)\left(1-e^{-t/\tau}\right)^{-1}, 
\end{equation}
where $\tau$ is the isotope's mean lifetime, and $t$ is the measurement time.

The production rates at Jungfraujoch $A_{\mathrm{J}}$ are calculated from the specific activities at the start of the measurement $A_{0}$, taking into account the activation time~$t_{\mathrm{a}}$ and the cool-down time~$t_{c}$. For short-lived nuclides, we must also take into account the rather short time interval $t_{\mathrm{t}}$, where the samples were exposed to a reduced cosmic ray flux during storage and transportation at lower altitudes, leading to a lower activation rate $A_{\mathrm{t}}$. The corrected specific activity $A_{\mathrm{J}}$ is
\begin{equation}
 A_{\mathrm{J}} = A_{0} e^{ \left( \frac{t_{\mathrm{c}}+t_{\mathrm{t}}}{\tau} \right)} \left[ \left( 1-e^{-t_{\mathrm{a}}/\tau} \right) + r\left( e^{t_{\mathrm{t}}/\tau}-1 \right) \right]^{-1}, 
\end{equation}
where $r=A_{\mathrm{t}}/A_{\mathrm{J}}$ is the ratio of vertical nucleon fluxes at Lauterbrunnen and the Jungfraujoch. We combine the storage and transport time to $t_{\mathrm{t}}=5$\,days and assume an altitude of 795\,m for the whole period. 
The specific activity of a sample of mass $m$ after an activation time $t_a$ is
\begin{equation}
A(t_a) = \frac{ N(t_a) }{ m \ \tau } = P \left( 1 - e^{-t_a/\tau} \right).
\end{equation}
The specific production (activation) rate $P$ is equal to the specific saturation activity
\begin{equation}
 A^0_{sat} = A(t_a \to \infty) = P
\end{equation}
for long activation times. In order to compare with other measurements and calculations, we convert $A_{\mathrm{J}}$ to the specific saturation activity at sea level $A_{\mathrm{sat}}$ by scaling it with the factor given in Table~\ref{tab:VertFlux}: 
\begin{equation}
A_{\mathrm{sat}} = A_{\mathrm{J}} / 11.2.
\end{equation}

\section{Results \label{Sec:Results}}

\subsection{Copper}

\begin{table*}[tbp]
\setlength\extrarowheight{4pt}
\centering
\caption{Results for the specific saturation activity $A_{\mathrm{sat}}$ of natural copper at sea level, derived from our measurements of the cosmogenic activation. These are compared to our predictions from Activia and Cosmo, using semi-empirical formulae for the cross sections. We also compare to a measurement performed at LNGS~\cite{Laub2009} (scaled to sea level by a factor 2.1 and corrected for an evaluation error in case of $^{54}$Mn and $^{59}$Fe~\cite{LaubPrivCom2015}), to predictions with Activia~\cite{ACTIVIA} using ($a$) the same semi-empirical formulae and ($b$) the MENDL-2P database for the cross sections, to semi-analytical calculations~\cite{Cebrian2010} using cosmic ray spectra from ($c$) Ziegler and ($d$) Gordon {\it et al.}, and to predictions using TALYS~\cite{Mei2009}. Deviations from our measured values beyond $+$1$\sigma$ and $-$1$\sigma$ are indicated by {\bf bold} or {\it italic} font styles, respectively. \label{tab:CuRes}}
\vspace{0.2cm}
\begin{tabular}{lrrrrrrrrrrr}
\hline
Isotope & T$_{1/2}$ & \multicolumn{9}{c}{Copper: specific saturation activity at sea level $A_{\mathrm{sat}}$ [$\mu$Bq/kg]} \\ [2pt]
\cline{3-12}
			& [days]  & \multicolumn{3}{c}{This work} 		&		& \multicolumn{6}{c}{Literature values} \\ [2pt]
\cline{3-5} \cline{7-12}
 & 
 & Measurement
 & \multicolumn{2}{c}{Calculations}
 & ~ 
 & Measurement
 & \multicolumn{2}{c}{Activia~\cite{ACTIVIA}}
 & \multicolumn{2}{c}{Calc.~\cite{Cebrian2010}}
 & Calculation
\\ 
 & 
 & 
 & Activia
 & Cosmo
 &  
 & LNGS~\cite{Laub2009}   
 & ${a}$
 & ${b}$
 & ${c}$
 & ${d}$
 & TALYS~\cite{Mei2009}  
\\ [2pt]
\hline
$^{46}$Sc 
 & $83.79$ 
 & $27^{+11}_{-9}$ 
 & $36$~~~ 
 & $17$~~~ 
 & ~ 
 & $25.2\pm8.6$ 
 & $36$ 
 & $36$ 
 & \bf{44} 
 & $31$ 
 & -- 
\\
$^{48}$V 
 & 15.97 
 & $39^{+19}_{-15}$ 
 & $34$~~~ 
 & $36$~~~ 
 & ~ 
 & $52\pm19$~ 
 & -- 
 & -- 
 & -- 
 & -- 
 & -- 
\\
$^{54}$Mn 
 & 312.12 
 & $154^{+35}_{-34}$ 
 & $166$~~~ 
 & $156$~~~ 
 & ~ 
 & $\mathbf{394\pm39}$~ 
 & $166$ 
 & $145$ 
 & {\bf 376} 
 & {\bf 321} 
 & $188$ 
\\
$^{59}$Fe 
 & 44.50 
 & $47^{+16}_{-14}$ 
 & $49$~~~ 
 & $50$~~~ 
 & ~ 
 & $57\pm14$~ 
 & ~~$49$ 
 & ~~{\it 21} 
 & ~~{\bf 75} 
 & ~~$57$ 
 & -- 
\\
$^{56}$Co 
 & 77.24 
 & $108^{+14}_{-16}$ 
 & $101$~~~ 
 & {\it 81}~~~ 
 & ~ 
 & $110\pm14$~ 
 & ~$101$ 
 & ~{\bf 163} 
 & ~{\bf 153} 
 & ~{\bf 231} 
 & -- 
\\
$^{57}$Co 
 & 271.74 
 & $519^{+100}_{\ -95}$ 
 & {\it 376}~~~ 
 & {\it 350}~~~ 
 & ~ 
 & $\mathbf{860\pm190}$ 
 & ~{\it 376} 
 & ~$421$ 
 & {\bf 1022} 
 & {\bf 858} 
 & {\bf 650} 
\\
$^{58}$Co 
 & 70.86 
 & $798^{+62}_{-58}$ 
 & {\it 656}~~~ 
 & {\it 632}~~~ 
 & ~ 
 & $786\pm43$~ 
 & {\it 655} 
 & ~{\it 441} 
 & {\bf 1840} 
 & {\bf 1430} 
 & -- 
\\
$^{60}$Co 
 & 1925.28 
 & $340^{+82}_{-68}$ 
 & $304$~~~ 
 & $297$~~~ 
 & ~ 
 & $\mathbf{1000\pm90}$~ 
 & $304$ 
 & {\it 112} 
 & {\bf 1130} 
 & ~{\bf 641} 
 & {\bf 537} 
\\ [2pt]

\hline
\end{tabular}
\end{table*}

The radioisotopes of interest were selected based on their half-life (T$_{1/2}\geq$5~days), the expected production rate, and on their $\gamma$-spectrum: we require at least one line at $E_\gamma \geq 60$\,keV with BR$\geq$10\%, which is separated from a prominent background line by more than 6\,$\sigma$.

Table~\ref{tab:CuRes} presents the results on the specific saturation activity at sea level $A_{\mathrm{sat}}$ for copper, derived from our activation sample. The numbers are compared with our own predictions using the Activia and Cosmo codes, with another measurement performed at LNGS~\cite{Laub2009}, and with additional predictions from the literature: these are based on the Activia package~\cite{ACTIVIA}, semi-analytical calculations~\cite{Cebrian2010}, and the TALYS~1.0~code~\cite{Mei2009,TALYS}. Deviations from our measurement beyond the $\pm$1$\sigma$ level are indicated by bold (too high) or italic (too low) font styles.

All cosmogenic radionuclides identified in the activated copper sample are well-known: they are produced in spallation reactions from the stable isotopes $^{63}$Cu and $^{65}$Cu~\cite{ACTIVIA,Laub2009,Cebrian2010}. The overall agreement between our measurement and the Activia/Cosmo calculations is remarkable, with most of the predicted activities within the 68\% C.I. The highest specific saturation activity, about 0.8~mBq/kg, is measured for $^{58}$Co which has a half-life of 71\,d. This value is $\sim$20\% higher than the prediction. The only other isotope where we measure a higher saturation activity ($\sim$30\%) than predicted is $^{57}$Co. For most of the isotopes the calculations with Cosmo yield systematically lower activities ($\sim$10\%), with the exception of $^{48}$V ($1.5\times$ higher than Activia) and $^{54}$Mn ($2\times$ higher). The general good agreement between measurement and predictions indicates the validity of our implementation of the cosmic ray flux at different altitudes, and can be considered as a benchmark for the comparison of the measurement and predictions for the xenon sample. 

For five out of eight isotopes, our results for copper agree with the only other measurement available in the literature, which has been performed at LNGS~\cite{Laub2009}. For $^{54}$Mn, $^{57}$Co and $^{60}$Co, we observe production rates which are $(2.5 \pm 0.6)$, $(1.7 \pm 0.5)$, and $(2.9 \pm 0.8)$ times lower, respectively. We obtain identical results with the Activia calculations in~\cite{ACTIVIA}, when we use the same semi-empirical formulae to calculate the excitation functions (case $a$). The predictions using the MENDL-2P data\-base~\cite{MENDL} (case $b$) tend to underpredict the production rates. The semi-analytical study~\cite{Cebrian2010} predicts much higher production rates than observed for both tested cosmic ray spectra (cases $c$~\cite{CRspectrum_Ziegler}, $d$~\cite{CRspectrum_Gordon}). The TALYS-based work~\cite{Mei2009} yields reasonable values, which are 20-60\% higher than measured, and show better agreement with our measurement than with the one of Ref.~\cite{Laub2009}.

\subsection{Xenon}

\begin{table*}[tbp]
\setlength\extrarowheight{3pt}
\centering
\caption{Results for the cosmogenic activation of natural xenon. The specific saturation activities at sea level are compared to our predictions based on Activia and Cosmo, to a measurement by LUX~\cite{LUXbg} and to a calculation using the TALYS code~\cite{Mei2009}. The half-lifes refer to the numbers used in the software packages. Deviations from our measured values beyond $+$1$\sigma$ and $-$1$\sigma$ are indicated by {\bf bold} or {\it italic} font styles, respectively.\label{tab:XeRes}}
\vspace{0.2cm}
\begin{tabular}{rrrrrcrr}
\hline
Isotope  & T$_{1/2}$ & \multicolumn{6}{c}{Xenon: specific saturation activity at sea level $A_\mathrm{sat}$ [$\mu$Bq/kg]} \\ [2pt]
\cline{3-8} & [days] & \multicolumn{3}{c}{This work} & & \multicolumn{2}{c}{Literature values} \\ [2pt]
\cline{3-5} \cline{7-8}
 & 
 & Measurement 
 &  \multicolumn{2}{c}{Calculations}
 & ~ 
 & Measurement
 & Calculation
\\ 
 & 
 & 
 & Activia 
 & Cosmo
 &  
 & LUX~\cite{LUXbg}  
 & TALYS~\cite{Mei2009}  
\\ [2pt]
\hline
$^{7}$Be~~~ 
 & 53.3 
 & $370^{+240}_{-230}$ 
 & {\it 6.4} 
 & {\it 6.4} 
 & ~ 
 & -- 
 & -- 
\\
$^{85}$Sr~~~ 
 & 64.8 
 & $<34$ 
 & $5.3$ 
 & $4.6$ 
 & ~ 
 & -- 
 & -- 
\\
$^{88}$Zr~~~ 
 & 83.4 
 & $<52$ 
 & $6.7$ 
 & $4.6$ 
 & ~ 
 & -- 
 & -- 
\\
$^{91\mathrm{m}}$Nb~~~ 
 & 62.0 
 & $<1200$ 
 & $5.6$ 
 & $5.0$ 
 & ~ 
 & -- 
 & -- 
\\
$^{99}$Rh~~~ 
 & 15.0 
 & $<120$ 
 & $8.3$ 
 & $8.2$ 
 & ~ 
 & -- 
 & -- 
\\
$^{101}$Rh~~~ 
 & 1205.3 
 & $1420^{+970}_{-850}$ 
 & {\it 16.6} 
 & {\it 15.3} 
 & ~ 
 & -- 
 & {\it 0.5} 
\\
$^{110\mathrm{m}}$Ag~~~ 
 & 252.0 
 & $<49$ 
 & $0.9$ 
 & $0.8$ 
 & ~ 
 & -- 
 & -- 
\\
$^{113}$Sn~~~ 
 & 115.0 
 & $<55$ 
 & $51$ 
 & $47$ 
 & ~ 
 & -- 
 & -- 
\\
$^{125}$Sb~~~ 
 & 986.0 
 & $590^{+260}_{-230}$ 
 & {\it 0.2} 
 & {\it 13.5} 
 & ~ 
 & -- 
 & {\it 0.5} 
\\
$^{121\mathrm{m}}$Te~~~ 
 & 154.0 
 & $<1200$ 
 & $299$ 
 & $276$ 
 & ~ 
 & -- 
 & $135$ 
\\
$^{123\mathrm{m}}$Te~~~ 
 &  119.7 
 & $<610$ 
 & $14.7$ 
 & $14.4$ 
 & ~ 
 & -- 
 & $140$ 
\\
$^{126}$I~~~ 
 & 13.0 
 & $175^{+94}_{-87}$ 
 & $247$ 
 & $247$ 
 & ~ 
 & -- 
 & -- 
\\
$^{131}$I~~~ 
 & 8.04 
 & $<190$ 
 & $147$ 
 & $170$ 
 & ~ 
 & -- 
 & -- 
\\
$^{127}$Xe~~~ 
 & 36.4 
 & $1870^{+290}_{-270}$ 
 & {\it 415} 
 & {\it 555} 
 & ~ 
 & $1530\pm300$  
 & -- 
\\
$^{129\mathrm{m}}$Xe~~~ 
 & 8.89 
 & $<8.7\times10^{3}$ 
 & $238$ 
 & $421$ 
 & ~ 
 & $1360\pm250$ 
 & -- 
\\
$^{131\mathrm{m}}$Xe~~~ 
 & 11.77 
 & $<3.6\times10^{4}$ 
 & $251$ 
 & $313$ 
 & ~ 
 & $1620\pm370$ 
 & -- 
\\
$^{133}$Xe~~~ 
 & 5.25 
 & $<1.2\times10^{5}$ 
 & $159$ 
 & $196$ 
 & ~ 
 & $1140\pm230$ 
 & -- 
\\
$^{132}$Cs~~~ 
 & 6.47 
 & $<120$ 
 & {\bf 166} 
 & {\bf 164} 
 & ~ 
 & -- 
 & -- 
\\ [2pt]
 
\hline
\end{tabular}
\end{table*}

For the xenon sample, we present the measured saturation activities at sea level in Table~\ref{tab:XeRes}, together with our Activia/Cosmo-based predictions. We also compare the results to measurement performed by the LUX Collaboration~\cite{LUXbg}, and to predictions using the TALYS code~\cite{Mei2009}. To re-scale the LUX-numbers to saturation activities at sea level, we use the procedure described in Section~\ref{Sec:Activation} and the information provided in Ref.~\cite{LUXbg}: we assume that all xenon was activated at sea level for 150~days, followed by 49\,days (7\,days) activation of 50\% (50\%) of the inventory at the SURF aboveground laboratory. The subsequent cool-down time underground was 90\,days (132\,days). The atmospheric depth at the SURF altitude of 1600\,m is 881\,g/cm$^{2}$, corresponding to a vertical nucleon flux of 8.5\,m$^{-2}$s$^{-1}$sr$^{-1}$. This yields a conversion factor of~3.3, in agreement with the number given by LUX~\cite{LUXbg}.

After activation, we have detected $\gamma$-lines from the following isotopes: $^7$Be, $^{101}$Rh, $^{125}$Sb,  $^{126}$I and $^{127}$Xe. While our measurement and predictions for $^{126}$I agree within the statistical and systematic errors, this is not the case for the light isotope $^7$Be, where the measurement is $\sim$50~times higher than the prediction, and for $^{125}$Sb, where we observe a ($2900 \pm 1200$)~times higher activity than predicted by Activia and a ($44 \pm 17$)~times higher activity than the Cosmo prediction. The observed production rate of $^{127}$Xe agrees with the measurement in the LUX detector~\cite{LUXbg}, however, the predictions are about a factor~4 too low. Our sensitivity did not allow us to detect the short-lived xenon isotopes $^{129\mathrm{m}}$Xe, $^{131\mathrm{m}}$Xe, and $^{133}$Xe. This also holds for the various other isotopes predicted by Activia and Cosmo, or by the study using TALYS ~\cite{Mei2009}. We note that the cosmogenic production rates for xenon isotopes predicted by Cosmo are systematically higher than the ones from Activia.

\section{Discussion and Conclusions }
\label{Sec:Conclusions}

We have carried out the first experiment to directly study the cosmogenic activation of xenon, which is employed as target and detection medium for rare event searches, and the second dedicated measurement on the activation of OFHC copper, often used as ultra-pure detector construction material. Both samples were activated by cosmic rays in a controlled manner at the Jungfraujoch research station (3470~m) to maximize the incident nucleon flux. The activation products were measured with a low-background germanium detector in the Gran Sasso underground laboratory. We have compared the measurement results to our own predictions obtained with the Activia~\cite{ACTIVIA} and Cosmo~\cite{COSMO} software packages, as well as to other measurements and predictions found in the literature.

From the eight detected isotopes in the OFHC copper sample, the production rates of five agree with a measurement performed at LNGS in 2009~\cite{Laub2009}, which is the only dedicated measurement  available in the literature. We measured up to a factor of $\sim$3~times lower values for the other three isotopes, see Table~\ref{tab:CuRes}. While the agreement with our Activia/Cosmo predictions is  satisfactory, supporting the validity of our comic ray model for the xenon study, discrepancies with other published calculations are present.

After cosmic activation of the xenon sample, we detected five isotopes, four of which were not directly measured before: $^{7}$Be, $^{101}$Rh, $^{125}$Sb, $^{126}$I. The measured saturation activity of $^{127}$Xe agrees with the measurement by the LUX collaboration~\cite{LUXbg}. In general, the agreement with predictions from Activia and Cosmos is unsatisfactory, the production rate for only one isotope, $^{126}$I, is calculated correctly. The predicted rates for most isotopes are too low by a factor of a few to $\sim$10$^3$, which is potentially worrisome, as some of these have rather long half-lives (3\,years). This conclusion also holds for the short-lived xenon isotopes $^{129\mathrm{m}}$Xe, $^{131\mathrm{m}}$Xe and $^{133}$Xe which we could not detect in our study: all predicted numbers are significantly lower than the values measured by LUX~\cite{LUXbg}. 

The main sources of background in multi-ton scale LXe-based dark matter detectors are the ones which are uniformly distributed in the target, such as $^{222}$Rn and $^{85}$Kr, or interactions from solar neutrinos~\cite{DARWIN}. Among the observed cosmogenic activation products of xenon, only $^{101}$Rh and $^{125}$Sb have half-lives which are long enough to affect an experiment with a foreseen operation time-scale of $\sim$5\,years. Both elements have a relatively high electronegativity and could possibly be removed by the xenon purification systems that use hot zirconium getters. However, this has not yet been demonstrated experimentally. The decay of $^{101}$Rh does neither produce low-energetic electrons nor low-energetic $\gamma$-rays not accompanied by a prompt ($<$1\,ns) second signal, which shifts the initial low energy signal to high energies by pile-up. We therefore conclude that $^{101}$Rh will not lead to dangerous background. The situation is different for $^{125}$Sb, where 13.6\% of the beta-decays end up in a long-lived excited state of the $^{125}$Te daughter ($T_{1/2}=57.4$\,d) and are therefore unaccompanied by a subsequent $\gamma$-ray. The low-energy tail of these electrons will lead to single-scatter electronic recoil background, while all other decay paths do not contribute, in agreement with the Geant4 prediction. Assuming that Sb is not removed by the purification system, and using the activation times quoted by LUX~\cite{LUXbg} together with our measured activation rate, a background contribution of $(3.0 \pm 1.3)\cdot 10^{-3}$\,events\,$\cdot$\,keV$^{-1}$\,$\cdot$\,kg$^{-1}$\,$\cdot$\,d$^{-1}$ would be expected. This rate is too large compared to the published total background level of $(3.6 \pm 0.4)\cdot 10^{-3}$\,events\,$\cdot$\,keV$^{-1}$\,$\cdot$\,kg$^{-1}$\,$\cdot$\,d$^{-1}$~\cite{LUXbg}. We therefore conclude that the true activation rate is either close to the lower end of our quoted credibility interval, or Sb is removed by the getter or plates out at surfaces.

In addition, the chemically inert noble gas isotopes could also affect next-generation dark matter searches. We have only observed $^{127}$Xe, which has a relatively short half-life ($T_{1/2}=36.3$\,d), similarly to other cosmogenically produced Xe-isotopes found in~\cite{LUXbg}. These isotopes will not pose a problem for next-generation LXe detectors, as their contamination will be reduced by an order of magnitude just after a few months of storage below ground. However, we note that for multi-ton scale LXe detectors such as LZ~\cite{ref::lz}, XENONnT~\cite{ref::xent} and eventually DARWIN~\cite{DARWINwimp,DARWIN}, the cosmogenic production of radioactive isotopes by the muon flux present at the underground location must be studied as well, as in this case also the decays of the short-lived isotopes might provide a source of backgrounds.

\section*{Acknowledgements}
We gratefully acknowledge support from the staff and the International Foundation High Altitude Research Stations Jungfraujoch and Gornergrat (HFSJG) operating the Jungfraujoch laboratory~\cite{Jungfrau}, where the activation with cosmic rays was performed, the SNF grants 200020-149256 and 20AS21-136660, the ITN Invisibles (Marie Curie Actions, PITN-GA-2011-289442), Dr.~D. Coderre,  L.~B\"utikofer, and Dr.~A.D.~Ferella for help with transportation and handling the samples at LNGS, and Prof.~Dr.~C.J.~Martoff for discussions on cosmogenic activation.

\end{document}